\documentclass[journal,twoside,web]{ieeecolor}
\usepackage{tmi}
\usepackage{cite}
\usepackage{amsmath,amssymb,amsfonts}
\usepackage{algorithmic}
\usepackage{graphicx}
\usepackage{textcomp}
\usepackage{subfigure}
\usepackage{bm}
\usepackage{multirow}
\let\labelindent\relax
\usepackage{enumitem}

\def\BibTeX{{\rm B\kern-.05em{\sc i\kern-.025em b}\kern-.08em
    T\kern-.1667em\lower.7ex\hbox{E}\kern-.125emX}}
\begin{document}
\title{Two-Stream Compare and Contrast Network for Vertebral Compression Fracture Diagnosis}
\author{Shixiang Feng, Beibei Liu, Ya Zhang, Xiaoyun Zhang, Yuehua Li,
\thanks{S. Feng, Y. Zhang, X. Zhang are with the Cooperative Medianet Innovation Center, Shanghai Jiao Tong University. (Email: \{fengshixiang, ya\_zhang, xiaoyun.zhang\}@sjtu.edu.cn).}
\thanks{B. Liu, Y. Li are with the Institute of Diagnostic and Interventional Radiology, Shanghai Sixth People's Hospital Affiliated to Shanghai Jiao Tong University (Email: \{beibei4906, liyuehua312\}@163.com).}}

\maketitle

\begin{abstract}
Differentiating Vertebral Compression Fractures (VCFs) associated with trauma and osteoporosis (benign VCFs) or those caused by metastatic cancer (malignant VCFs) are critically important for treatment decisions. So far, automatic VCFs diagnosis is solved in a two-step manner, i.e. first identify VCFs and then classify it into benign or malignant. In this paper, we explore to model VCFs diagnosis as a three-class classification problem, i.e. normal vertebrae, benign VCFs, and malignant VCFs. However, VCFs recognition and classification require very different features, and both tasks are characterized by high intra-class variation and high inter-class similarity. Moreover, the dataset is extremely class-imbalanced. To address the above challenges, we propose a novel Two-Stream Compare and Contrast Network (TSCCN) for VCFs diagnosis. This network consists of two streams, a recognition stream which learns to identify VCFs through comparing and contrasting between adjacent vertebra, and a classification stream which compares and contrasts between intra-class and inter-class to learn features for fine-grained classification. The two streams are integrated via a learnable weight control module which adaptively sets their contribution. The TSCCN is evaluated on a dataset consisting of 239 VCFs patients and achieves  the average sensitivity and specificity of 92.56\% and 96.29\%, respectively.
\end{abstract}


\section{Introduction}
\IEEEPARstart{V}{ertebral} compression fractures (VCFs) are associated with back pain, disability, and limitation of spine mobility which leads to health-related deterioration of quality of life \cite{szulc2011overview,oei2013review}.  The causes of VCFs include trauma, osteoporosis, and neoplastic infiltration. VCFs caused by trauma and osteoporosis are generally benign, whose incidence rate increases with age. Certain cancers, such as thyroid and breast which have a propensity to metastasize to bone, can lead to malignant VCFs \cite{mauch2018review}. Differentiating between benign and malignant VCFs is critically important for the treatment and prognostic of patients. MRI is considered to be the most reliable imaging method for spine diseases \cite{uetani2004malignant}. Signal intensity and shape of vertebral bodies are important imaging features when diagnosing VCFs in MRI \cite{kalid2005review,cuenod1996acute}. Spectral and fractal features \cite{azevedo2015classification} that are related to signal intensity, and compactness and convex deficiency features \cite{frighetto2015classification} that are related to shape, were widely employed for VCFs diagnosis.

\begin{figure}[!t]
\begin{center}
    \subfigure[]{
    \includegraphics[width=0.21\linewidth]{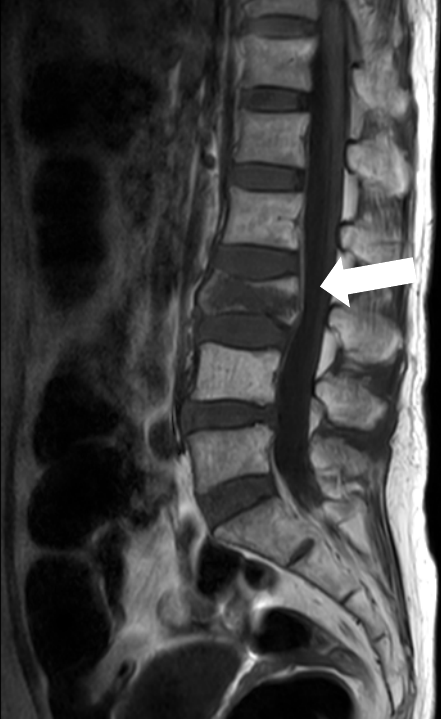}
  }
    \subfigure[]{
    \includegraphics[width=0.21\linewidth]{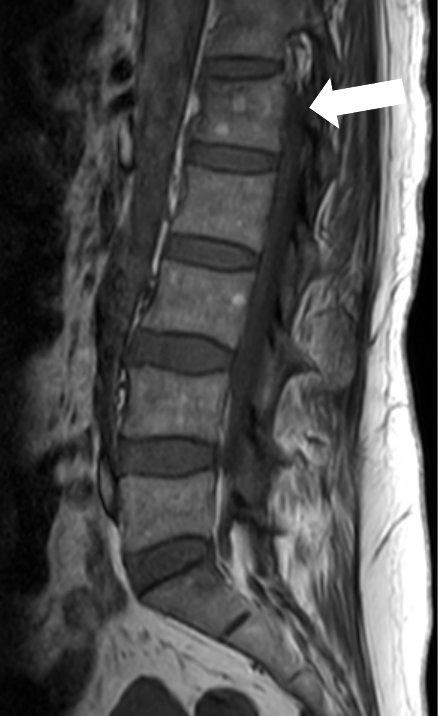}
    }
    \subfigure[]{
    \includegraphics[width=0.21\linewidth]{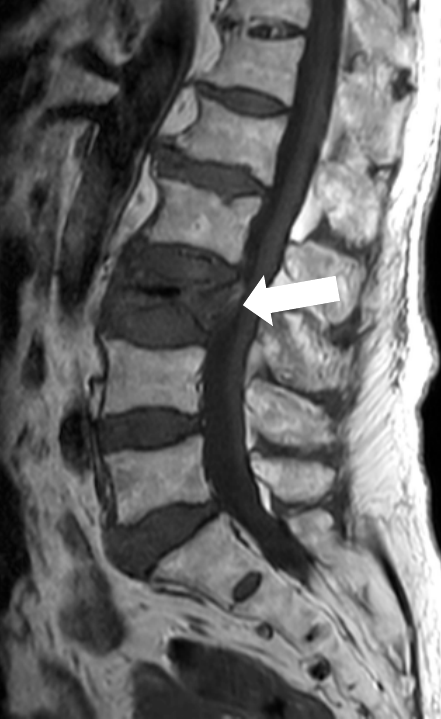}
  }
    \subfigure[]{
    \includegraphics[width=0.21\linewidth]{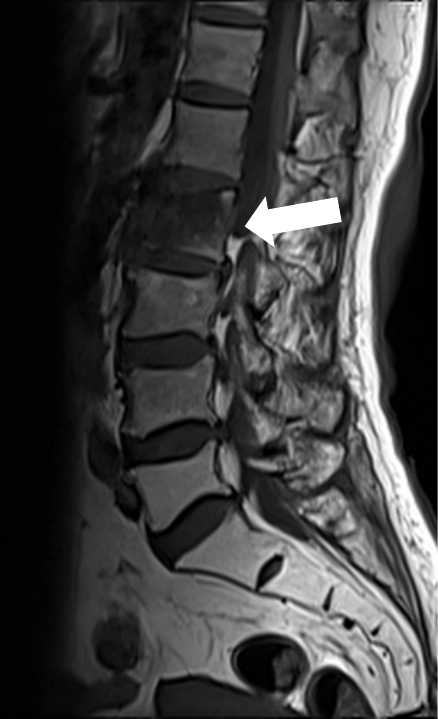}
    }
\end{center}
   \caption{Examples of spine T1-weighted MRI used in the study. The arrows point to the VCFs. When only focusing on one vertebra, (a) the VCFs can be easily identified, (b) the VCFs may be diagnosed as a normal vertebra, (c) and (d) some normal vertebrae may be wrongly diagnosed as VCFs since the bad spine health condition of the patients. Utilizing the continuity of the vertebrae by comparing the current vertebra to its adjacent vertebrae may help improve diagnosis accuracy.}
\label{img2}
\end{figure}

\begin{figure}[!t]
\centerline{\includegraphics[width=\columnwidth]{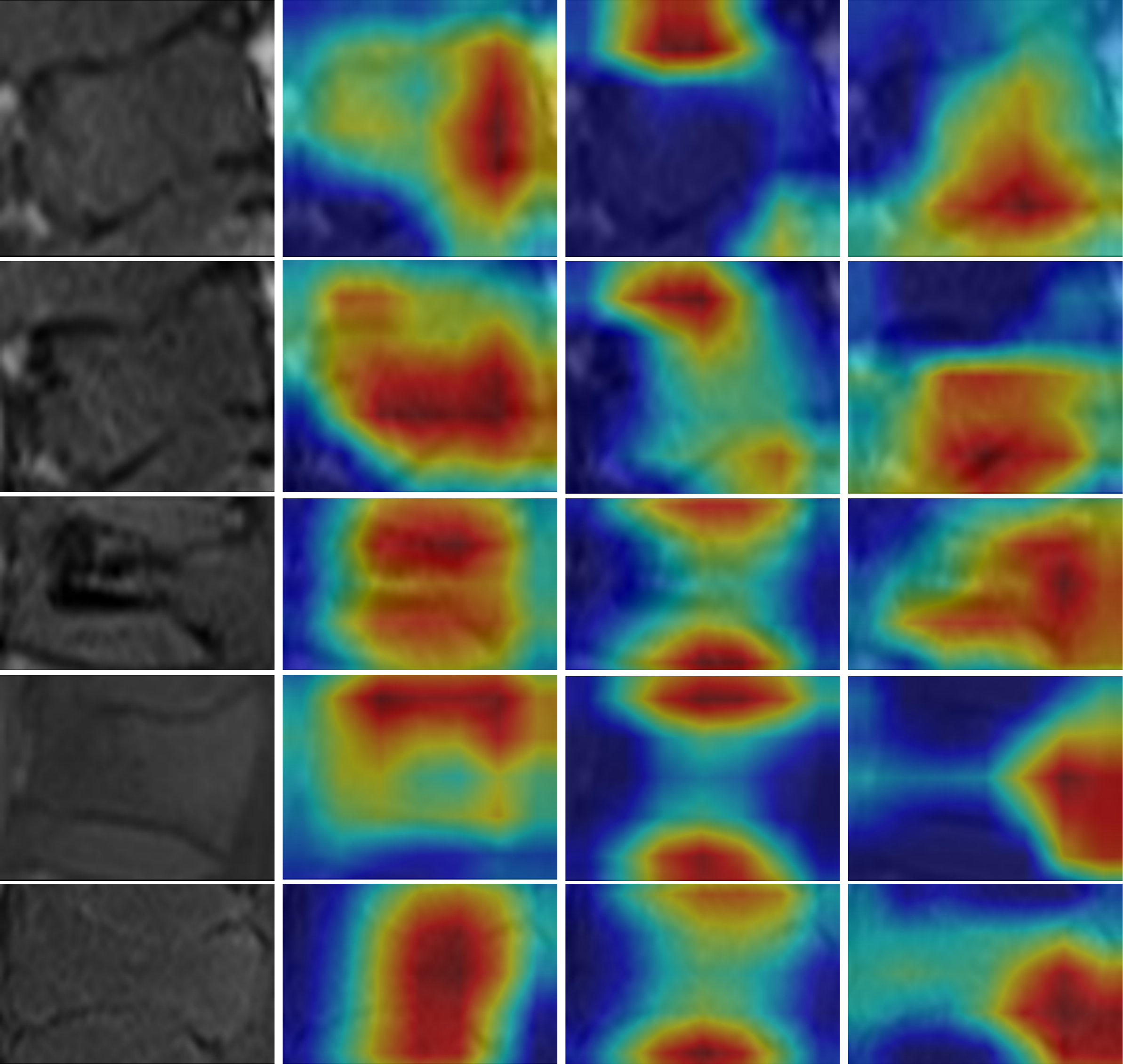}}
\caption{Examples of CAM images. ResNet is used for different classification tasks: three-class, normal-fractured, and benign-malignant classification. The first column shows vertebral bodies. The second, third, and last columns show CAM images of three-class classification, normal-fractured classification, and benign-malignant classification, respectively.}

\label{cam}
\end{figure}

Typically, VCFs diagnosis is divided into recognition, i.e. differentiating between normal vertebrae and VCFs, and classification, i.e. differentiating between benign and malignant VCFs.
Most of the previous studies only solve one of the tasks \cite{bromiley2017classification, bar2017compression,tomita2018deep, frighetto2015semiautomatic,azevedo2015classification}. Frighetto \textit{et al.} \cite{frighetto2016recognition} solve both two tasks but in a two-step manner, i.e. VCFs are recognized first and then classified based on the recognition results. The two-step solution is expected to lead to a problem that the performance of the first step directly affects that of the second step. Classification of the three classes in a single step appears to be feasible \cite{frighetto2016recognition} and can avoid this problem. 
However, performing VCFs diagnosis in a one-step manner faces three challenges.
The first challenge is that VCFs recognition and VCFs classification require different features. As shown in Fig. \ref{cam}, for recognition, the network primarily focuses on the upper and lower edges of the vertebral bodies, while for classification, the network focuses on more local parts of the vertebral bodies. This subtle conflict implies a trade-off between these two tasks, which might reduce a single integrated network's diagnosis performance.
The second challenge is the intra-class variation and inter-class similarity in both VCFs recognition and classification tasks. As shown in Fig. \ref{img2}(b), slightly fractured vertebrae are similar to normal vertebrae. Furthermore, the shape and signal intensity vary within one subtype of VCFs, while are similar between subtypes of VCFs.
The third challenge is class imbalance. VCFs are much less than normal vertebrae and class imbalance is exacerbated since the VCFs are divided into benign and malignant. The class imbalance makes the network easily overfit to minority classes.
As a result, formulating the VCFs diagnosis as a naive classification problem is sub-optimal, more so in the case of fine-grained and imbalanced data regimes.

In this paper, we propose a Two-Stream Compare and Contrast Network (TSCCN) to address the above challenges. 
For the rest of the paper, we assume the vertebrae are already segmented and we directly worked on segmented individual vertebrae.
% which only needs four corner landmarks for training.
% The two-stream architecture has recently been demonstrated effectiveness for enhancing the extraction of different features respectively \cite{zhang2020texture, jin2019accurate}. In TSCCN, two streams are used to extract features related to VCFs recognition and VCFs classification separately to address the trade-off between recognition and classification.
TSCCN is composed of two streams, one focusing on recognition and the other for classification, so that respective feature representations are learned. 
To address the fine-grained nature of the recognition and classification tasks, each stream learns through \emph{compare and contrast}. 
Specifically, considering the high variation of vertebra appearance from upper spine to lower spine, the recognition stream uses the proposed Compare Network to leverage three input branches to learn by comparing the current vertebra to its adjacent vertebrae, utilizing the continuity of the vertebrae in the spine. The classification stream adopts the triplet loss to compare and contrast intra-class with inter-class. A benign-malignant classification head is further added to the stream so that it focuses more on distinguishing between benign and malignant.
An additional benefit of this two-stream architecture is to alleviate the impact of class imbalance. This class-imbalanced task is decomposed into (1) classifying between the majority class (normal vertebrae) and the collection of minority classes (VCFs) and (2) classifying within minority classes, so the influence of the majority class is weakened when classifying the subtypes of minority classes.
For a two-stream architecture, how to effectively integrate the features from two streams is critical. A naive solution is simple concatenation of the features, which has been widely practiced in previous studies \cite{zhang2020texture, jin2019accurate}.
In TSCCN, we propose a learnable weight control module for features integration to automatically control the contribution of two streams. Different weights are assigned to the features from different streams according to the prediction of the network.

The contribution of this paper is summarized as follows:
\begin{itemize}
    \item We propose a novel two-stream compare and contrast network to simultaneously perform VCFs recognition and classification. To our best knowledge, we are the first to solve the three-class classification task end-to-end. %The proposed network is composed of two streams to extract features for VCFs recognition and classification, respectively. 
    \item We leverage the compare and contrast among neighboring vertebrae to learn representation to distinguish between normal vertebrae and VCFs.
    \item To achieve fine-grained classification between benign and malignant VCFs, we introduce the triplet loss to compare between inter-class and intra-class. 
    \item To better integrate the features from two streams, we propose a novel two-stream integration method which gates the contribution of features from two streams with a learnable weight control module.
\end{itemize}

\section{Related Work}
\label{sec2}
\subsection{Vertebral Compression Fractures Diagnosis}
Vertebrae segmentation and localization attract lots of interests \cite{chen2019vertebrae,lessmann2019iterative,peng2019weakly} while only a few methods have been proposed for further analysis such as  VCFs recognition and VCFs classification.
Previous works extract hand-crafted features and conduct the recognition and classification tasks separately \cite{bromiley2017classification,azevedo2015classification,frighetto2015classification,frighetto2016recognition}. Bromiley \textit{et al.} \cite{bromiley2017classification} combine random forest classifiers and appearance modelling to do VCFs recognition task. Azevedo \textit{et al.} \cite{azevedo2015classification} extract spectral and fractal fractures from manually segmented images to do the task of VCFs classification. Frighetto \textit{et al.} \cite{frighetto2016recognition} conduct both two tasks but in a two-step manner. They detect the VCFs from the spine first and then classify the VCFs into benign and malignant, and the same method is used to extract features for the two tasks.
Bar \textit{et al.} \cite{bar2017compression} propose a CNN-based method for VCFs recognition in CT scans. They use a CNN to classify sagittal patches extracted from the vertebral column, and a recurrent neural network to aggregate the classification across patches. Similar to \cite{bar2017compression}, Tomita \textit{et al.} \cite{tomita2018deep} employ a CNN to extract features for sagittal CT slices, and these features are aggregated by using a long short-term memory network to make a diagnosis. Different from \cite{bar2017compression}, \cite{tomita2018deep} does not need segment the spine first, but it can not localize the VCFs. However, these methods are only simple implementations of CNN working on VCFs recognition.

\subsection{Fine-Grained Visual Classification in Medical Image}
With the development of neural networks, effective networks \cite{he2016deep,simonyan2014very} are proposed to extract representative features for general image classification tasks, but they do not perform well when they are applied for fine-grained visual classification (FGVC). 
% For nature images, the methods proposed in recent years can be roughly divided into two categories. The methods that belong to the first category focus on feature encoding. Bilinear-CNN 
% \cite{lin2015bilinear} uses a two-stream network architecture and computes the outer product of the outputs of the two streams using a bilinear pooling module to capture the second-order information.
% Gao \textit{et al.} \cite{gao2016compact} improve bilinear model by updating it into a compact structure and Kong \textit{et al.} \cite{kong2017low} use a bilinear classifier replacing the bilinear feature.
% Cui \textit{et al.} \cite{cui2017kernel} propose a kernel pooling method to capture higher-order representation.
% The methods belong to the second category focus on locating the discriminative parts and extracting features of these parts without using object/part annotation. Zhang \textit{et al.} \cite{zhang2016picking} propose a two-step approach to pick filters responding to specific parts and learn part detectors. Zheng \textit{et al.} \cite{zheng2017learning} jointly improve the performance of part localization and feature learning and train a fusion layer to fuse features of different parts. Ge \textit{et al.} \cite{ge2019weakly} focus on complementary object parts by fusing features from these parts using LSTMs.
For medical image classification tasks, classification of subtypes of deceases is a common problem. Fine-grained classification attracts many interests, since intra-class variation and inter-class similarity are common in many kinds of diseases, such as skin lesion classification \cite{zhang2019attention, zhang2018skin} and lung nodules classification \cite{shen2017multi}. The methods proposed to improve the features extraction ability of networks can be roughly divided into two categories.
The first category focuses on proposing new network architecture to better extract representative features. Zhang \textit{et al.} \cite{zhang2018skin} propose a synergic deep learning (SDL) model, which uses two networks with the same architecture and enables them to learn from each other. Zhang \textit{et al.} \cite{zhang2019attention} propose an attention residual learning (ARL) block which combines attention learning and residual learning to improve its extraction ability for discriminative features and their results show the ability of ARL-CNN to focus on discriminative parts of skin lesions. Shen \textit{et al.} \cite{shen2017multi} propose the multi-crop pooling strategy to capture nodule salient information. 
The methods that belong to the second category focus on utilizing expert knowledge. Lai \textit{et al.} \cite{lai2019spatial} utilize the expert knowledge that the spinal edges are important for dislocation diagnosis and they use the knowledge to guide model training by introducing a spatial regularization term. In this work, we combine these two categories. Expert knowledge is utilized to guide the design of our network. Compare network is proposed to emulate the knowledge about recognizing VCFs.

\subsection{Imbalanced Classification in Medical Image}
Imbalanced classification attracts lots of interest in researchers. The performance of networks designed for class-balanced classification tasks is not satisfying when they are applied to class-imbalanced classification tasks, because they will put emphasis on the majority classes and fail to classify the minority classes \cite{dong2018imbalanced,ren2018learning}. 

Data imbalance is common in medical data and many works have pointed out the problem in many tasks, such as vertebral fracture grading \cite{husseini2020grading}, skin lesion classification \cite{zhang2018skin}, and lung nodule classification \cite{eun2018single}. However, there are few works dedicated to solving this problem explicitly.
Sakamoto \textit{et al.} \cite{sakamoto2018lung} use a cascaded network and a fusion classifier to classify the lung nodule for a class imbalanced nodule candidate dataset. Two basic strategies popular in the natural image (a) over-sampling the minority classes \cite{japkowicz2000class, cui2018large} or down-sampling the majority classes \cite{he2009learning}, and (b) re-weighting the loss function in a class-wise manner, e.g., higher factors for minority classes \cite{huang2016learning}, are used to solve the imbalanced data problem.
Zhang \textit{et al.} \cite{zhang2018skin} control the proportion of input intra-class image pairs to avoid imbalance data problem.
Eun \textit{et al.} \cite{eun2018single} augment nodules by translation and rotation to address this problem. 
For natural images, Cui \textit{et al.} \cite{cui2019class} define the concept of "effective number" which denotes the volume of sample and is used to re-weight the loss. Cao \textit{et al.} \cite{cao2019learning} propose label-distribution-aware margin (LDAM) loss to optimize the generalization error bound and find that applying re-weighting or re-sampling at the later stage of training can be much more effective. However, the effectiveness of these methods proposed on natural images in recent years has not been verified in medical images. To address this problem we use a simple but effective method. Two minority classes are regarded as one class, VCFs, and classifying between minority classes is performed separately to avoid classifying minority classes with the existence of majority class.

\section{Methodology}
\label{sec3}
\begin{figure*}[!t]
\centerline{\includegraphics[width=2\columnwidth]{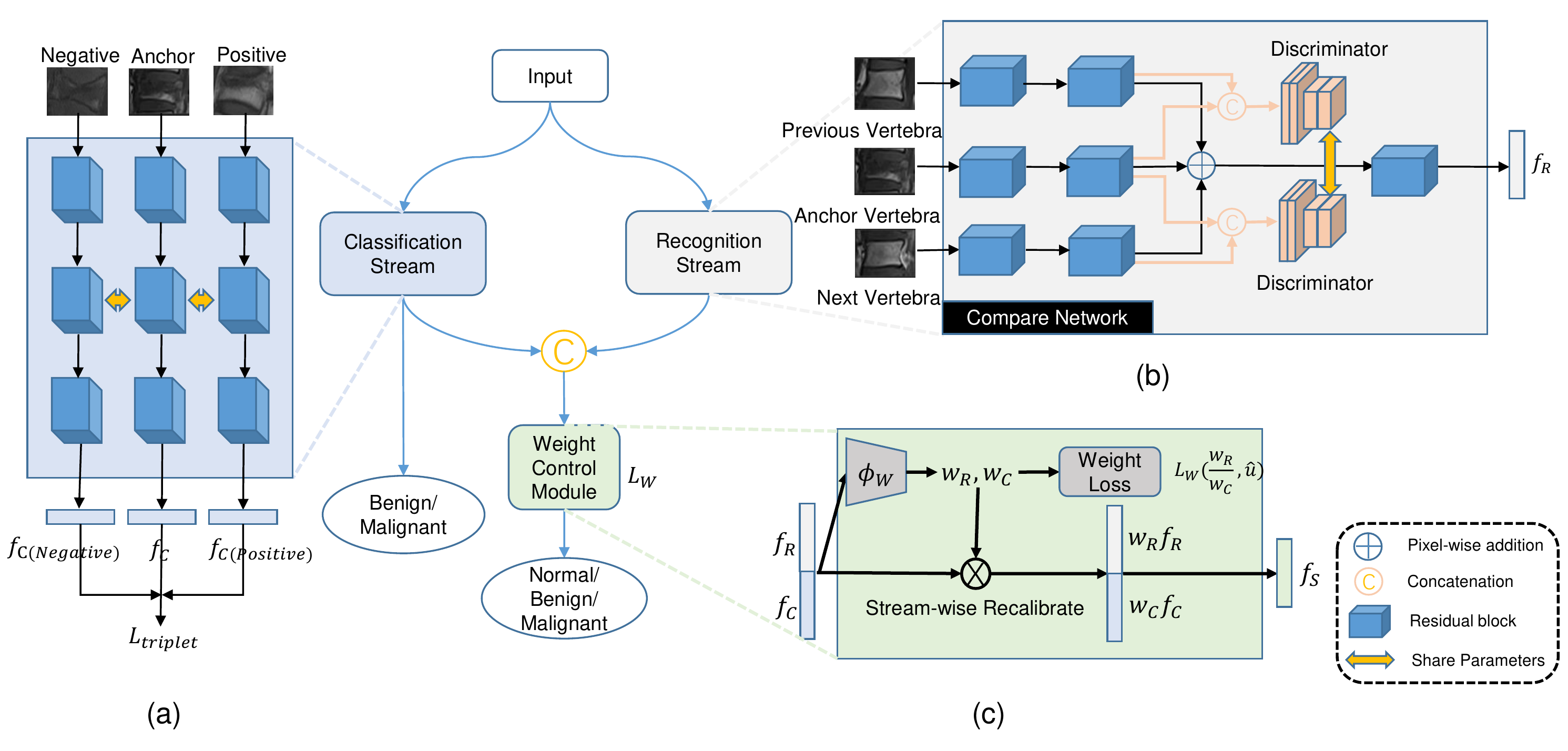}}
\caption{Network overview. (a) depicts the classification stream that compares and contrasts benign and malignant vertebrae, (b) depicts recognition stream which applies compare network to compare and contrast the current vertebra to adjacent vertebrae, (c) illustrates weight control module which explicitly controls the contribution of two streams and employs stream-wise recalibration.}
\label{arch}
\end{figure*}

The proposed Two-Stream Compare and Contrast Network (TSCNN) is depicted in Fig. \ref{arch}.  The input to TSCCN is segmented individual vertebrae. For now we assume the vertebrae is segmented and we defer the details on vertebrae segmentation to Sec \ref{seg}.
To address the challenge that recognition and classification require related but different features (Fig. \ref{cam}), TSCNN is composed of two streams, one for recognition and the other for classification.
The recognition stream applies the proposed Compare Network to leverage a three-branch architecture to compare and contrast adjacent vertebrae so as to identify VCFs from vertebrae.
The classification stream adopts a cross-entropy loss to differentiate benign and malignant VCFs. Considering the difference between the two types of VCFs are very fine-grained, a triplet loss is further introduced to compare and contrast them.
For the final three-class classification, a weight control module is proposed to explicitly control the contribution of the two streams during feature fusion.

\subsection{Recognition Stream}

As depicted in Fig. \ref{arch}(b), Compare Network is applied to recognition stream, which is composed of three input branches with identical structures and a discriminator.

\subsubsection{Three input branches}
Recognizing VCFs by focusing on only one vertebra is not always accurate. Experienced radiologists usually compare the current vertebra with its adjacent vertebrae, utilizing the continuity of the spine, i.e. the normal vertebra is similar in the shape and signal intensity to its adjacent normal vertebrae. 
Mimicking the radiologists, the compare network is designed to take three neighboring vertebrae as input, i.e. the current vertebra and its previous and next vertebrae (denote as $x_c,\,x_l,\,x_n$ respectively), and leverage a three-branch architecture to compare and contrast adjacent vertebrae to identify VCFs from vertebrae. We denote the three branches corresponding to $x_c,\, x_p,\,x_n$ as $f_c,\, f_p,\, f_n$ respectively. The three branches have the same network structure. The features obtained by the three branches are fused by pixel-wise addition, and then input to the rest convolution layers of the stream $M$. The final output features of this stream $f_{R}$ is expressed as:
\begin{equation}
f_{R} = M(f_l(x_p)+f_c(x_c)+f_n(x_n)).
\label{eq1}
\end{equation}

\subsubsection{Discriminator}
Inspired by the SDL network \cite{zhang2018skin}, after obtaining the output features of the three branches, the features of the current vertebra and its previous adjacent vertebra are concatenated and fed into a discriminator. We do the same to the current vertebra and its next vertebra. The discriminator is used to determine whether the two vertebrae belong to the same class.
The loss of the discriminator $D$ can be expressed as:
\begin{equation}
\begin{aligned}
L_D=&L_{CE}(D(f_c(x_c),f_l(x_l)), |y_c-y_l|)\\
+&L_{CE}(D(f_c(x_c),f_n(x_n)), |y_c-y_n|),
\end{aligned}
\label{eq2}
\end{equation}
where $y_i\in\{0,1\}, i\in\{c, l, n\}$ is the label of the current vertebra, previous vertebra and next vertebra, respectively, and $y_i=0$ if the vertebra is normal and $y_i=1$ for the benign and malignant VCFs. 

The discriminator compares and contrasts adjacent vertebrae for VCFs recognition.  
An additional benefit of the discriminator is to increase the contribution of the center vertebrae. By comparing between $x_c$ and $x_l$, $x_c$ and $x_n$ using discriminator, as shown in \eqref{eq2}, $x_c$ contributes more than $x_l$ and $x_n$, which is consistent with the idea of taking the center vertebra as the main judgment basis and the adjacent vertebrae as the auxiliary information for comparison.

\subsection{Classification Stream}

Considering the characteristic that VCFs have large variations in shape and signal intensity even for the same subtype of VCFs, we introduce triplet loss \cite{schroff2015facenet} to compare and contrast benign and malignant VCFs. Triplet loss compares the anchor vertebra to a positive vertebra and a negative vertebra, and the distance from the anchor to positive vertebrae is minimized while the distance from anchor to negative vertebrae is maximized. Thus, the representations of vertebrae of the same class are clustered together while those of different classes are pushed apart.  
A binary classification head is further added to classification stream. The auxiliary binary classification head emphasizes the benign-malignant biased features extraction ability. 
The loss of the classification stream is
\begin{equation}
L_{C}=L_{CE2}+L_{triplet},
\label{eq3}
\end{equation}
where $L_{CE2}$ denotes the cross-entropy loss for benign and malignant classification. The stream loss is calculated and back-propagated only when the input images are benign or malignant VCFs during training.

\subsection{Two Streams Integration}
The features from two streams are integrated to make the final three-class classification. Integration by simply concatenating makes each stream contribute equally, which may disregard highly informative features from a certain stream. In this work, we introduce a weight control module which  generate adaptive weight $w$ to integrate features which controls the contribution of two streams.
Denote the output features of the two streams as $f_{R}$ and $f_{C}$. As shown in Fig. \ref{arch}(c), $f_R$ and $f_c$ are first concatenated and a learnable function $\phi_W$ is employed to transform the concatenated features to weight $w=[w_R, w_c]$ with dimension 2:
\begin{equation}
w=\phi_W(f_R,f_C).
\label{eq4}
\end{equation}
Inspired by Squeeze-and-Excitation Network \cite{hu2018squeeze}, global average pooling and an MLP is applied for $\phi_W$. $w_R$ and $w_C$ determine the weights of the two features. The fused feature $f_S$ is obtained by concatenating $w_R*f_{R}$ and $w_C*f_{C}$ and is input to the three-class classification head to make the final prediction. 

When the input current vertebra is normal, $f_{R}$ should be more important than $f_{C}$, and vice versa. We propose the weight loss to explicitly control the weights according to the label of input vertebra. We denote weight ratio $u=\frac{w_R}{w_C}$, and we set a parameter $\hat{u},(\hat{u}>1)$, which controls the bound of $u$. The weight loss can be expressed as: 
\begin{equation}
\label{eq5}
L_W=\left\{
\begin{aligned}
&\|u-\hat{u}\|^2,&y=0,~u<\hat{u},or~y\in\{1,2\},~u>\frac{1}{\hat{u}}, \\
&0, & otherwise.
\end{aligned}
\right.
\end{equation}
In this work, we set $\hat{u}$ as 4.

Finally, the total loss of TSCCN is 
\begin{equation}
L=L_{CE}+\lambda_1L_D+\lambda_2L_{C}+\lambda_3L_W,
\label{eq6}
\end{equation}
where $L_{CE}$ is the three-class cross-entropy loss. $\lambda_1,\lambda_2$, and $\lambda_3$ are parameters weighting the importance of each component.

\subsection{Segmentation and Post-processing}
\label{seg}
The VCFs diagnosis is naturally a fine-grained recognition task. To model it as a typical classification problem, each vertebra is first cropped into a patch from the MRI slices according to coarse segmentation results. A weakly-supervised vertebrae segmentation method, WISS \cite{peng2019weakly}, which only needs four corner landmarks on a single sagittal slice, is used to segment MRI spine images. WISS fails to segment the seriously fractured vertebrae because of the severe collapse of vertebral bodies and diminished contrast to the surrounding structures. We design a post-processing method to get a better segmentation result. The process of the automatic post-processing is: (a) delete the small connected areas to remove the masks of under-segmented fractured vertebrae, (b) make up the mask for the fractured vertebra in (a) according to the distance between adjacent vertebrae and copy the mask from adjacent vertebra. By applying the post-processing method, almost all of the lost severely fractured vertebrae can be made up. For patients with continuous severely fractured vertebrae, which is rare in our datasets, making up all the lost vertebrae may fail. However, at least one vertebra can be made up, so the possible missing segmentation will not affect the final patient-level prediction.
It should be noticed that we do not need accurate segmentation masks but coarse segmentation masks, since margin is added when cropping to ensure the patches contain the whole vertebral bodies.

\begin{table*}
\caption{Results of competing methods and ours. For methods belonging to fine-grained classification (PC-Net, NTS-Net, SDL-Net), over-sampling the minority classes is applied. 0, 1, 2 are used to denote normal vertebrae, benign VCFs, and malignant VCFs, respectively. The best and second best results for each column are marked in bold and with underline, respectively.}
\setlength{\tabcolsep}{3pt}
\center
\begin{tabular}{lcccccc|cccccc}
\hline
Method & $SE_0$ & $SP_0$& $SE_1$ & $SP_1$ & $SE_2$ & $SP_2$ & aSE(\%) & aSP(\%)& aAUC(\%) & mAP(\%) \\
\hline
\hline
ResNet\cite{he2016deep} & $95.45_{\pm0.96}$ & $92.41_{\pm2.00}$ & $86.89_{\pm5.01}$ & $87.01_{\pm3.22}$ & $66.57_{\pm5.89}$ & $95.09_{\pm0.94}$ & $82.97_{\pm1.49}$ & $91.50_{\pm0.80}$ & $95.42_{\pm1.17}$ & $91.34_{\pm2.67}$ \\
\hline
RS\cite{japkowicz2000class} & $94.90_{\pm2.65}$ & $93.77_{\pm2.46}$ & $84.00_{\pm6.01}$ & $90.85_{\pm4.04}$ & $76.57_{\pm4.36}$ & $93.92_{\pm2.92}$ & $85.56_{\pm1.54}$ & $92.85_{\pm0.74}$ & $96.33_{\pm0.27}$ & $94.13_{\pm0.58}$  \\
CB-RW\cite{cui2019class} & $95.82_{\pm1.03}$ & $93.34_{\pm1.28}$ & $87.55_{\pm5.35}$ & $91.19_{\pm1.74}$ & $75.71_{\pm5.63}$ & $95.08_{\pm2.85}$ & $86.36_{\pm1.34}$ & $93.20_{\pm0.66}$ & $96.51_{\pm0.86}$ & $93.70_{\pm1.68}$  \\
SMOTE\cite{chawla2002smote} & $93.92_{\pm1.72}$ & $94.14_{\pm1.44}$ & $90.00_{\pm1.57}$ & $91.52_{\pm1.58}$ & $79.29_{\pm0.83}$ & $95.97_{\pm0.54}$ & $87.74_{\pm0.36}$ & $93.88_{\pm0.19}$ & $96.98_{\pm0.42}$ & $94.53_{\pm0.96}$ \\
LDAM\cite{cao2019learning} & $96.63_{\pm1.27}$ & $93.10_{\pm2.31}$ & $\underline{91.11}_{\pm2.63}$ & $91.00_{\pm2.27}$ & $74.52_{\pm4.74}$ & $97.01_{\pm1.44}$ & $87.42_{\pm1.75}$ & $93.70_{\pm0.87}$ & $94.43_{\pm1.86}$ & $90.65_{\pm3.00}$  \\
LDAM+DRW\cite{cao2019learning} & $92.52_{\pm3.08}$ & $\pmb{96.47}_{\pm1.21}$ & $86.00_{\pm4.42}$ & $94.32_{\pm3.01}$ & $\underline{88.29}_{\pm3.40}$ & $92.78_{\pm1.70}$ & $88.94_{\pm1.39}$ & $94.53_{\pm0.74}$ & $96.76_{\pm0.37}$ & $93.92_{\pm1.08}$ \\
\hline
PC-Net\cite{dubey2018pairwise} & $92.39_{\pm1.84}$ & $\underline{95.99}_{\pm1.08}$ & $84.58_{\pm2.56}$ & $\underline{95.31}_{\pm2.37}$ & $88.21_{\pm3.73}$ & $91.46_{\pm3.55}$ & $88.40_{\pm1.51}$ & $94.25_{\pm0.71}$ & $97.18_{\pm0.37}$ & $95.18_{\pm0.72}$\\
NTS-Net\cite{yang2018learning} & $\pmb{97.30}_{\pm0.87}$ & $91.63_{\pm1.76}$ & $88.50_{\pm3.41}$ & $93.78_{\pm1.07}$ & $82.70_{\pm2.50}$ & $\textbf{97.97}_{\pm1.09}$ & $89.32_{\pm0.99}$ & $94.64_{\pm0.48}$ & $\underline{98.01}_{\pm0.34}$ & $\underline{96.19}_{\pm0.78}$\\
SDL-Net\cite{zhang2018skin} & $\underline{97.19}_{\pm0.44}$ & $93.99_{\pm1.80}$ & $88.89_{\pm2.22}$ & $94.17_{\pm1.14}$ & $83.57_{\pm3.40}$ & $96.78_{\pm0.19}$ & $\underline{89.88}_{\pm1.56}$ & $\underline{94.98}_{\pm0.78}$ & $97.00_{\pm0.82}$ & $94.87_{\pm1.85}$ \\
\hline
Ours & $95.12_{\pm1.40}$ & $94.14_{\pm1.92}$ & $\pmb{91.85}_{\pm3.63}$ & $\pmb{96.85}_{\pm1.85}$ & $\pmb{90.71}_{\pm3.59}$ & $\underline{97.86}_{\pm1.45}$ & $\pmb{92.56}_{\pm1.46}$ & $\pmb{96.29}_{\pm0.75}$ & $\pmb{98.35}_{\pm0.49}$ & $\pmb{97.01}_{\pm0.88}$ \\
\hline
\end{tabular}
\label{tab2}
\end{table*}

\section{Experimental Results}
\label{sec4}
\subsection{Dataset and Evaluation metrics}
\subsubsection{Dataset}
We collect a dataset consisting of $239$ patients' T1-weighted MRI images with VCFs, including $136$ patients with benign VCFs and $103$ patients with malignant VCFs. 
%All the images were scanned with $1.5$ Tesla MR scanner with following protocol: Slice thickness $3.0$ mm, Pixel Spacing $1.25$ mm, Repetition Time (TR) = $10.6$ ms, Echo time (TE) = $4.76$ ms. 
For each patient, $10\sim13$ sagittal slices are provided. Patient-level labels of malignant and benign and coordinates of the VCFs are given. All the malignant patients are examined by pathological biopsy. All images are annotated and examined by experienced radiologists.

\begin{figure}[!t]
\centerline{\includegraphics[width=\columnwidth]{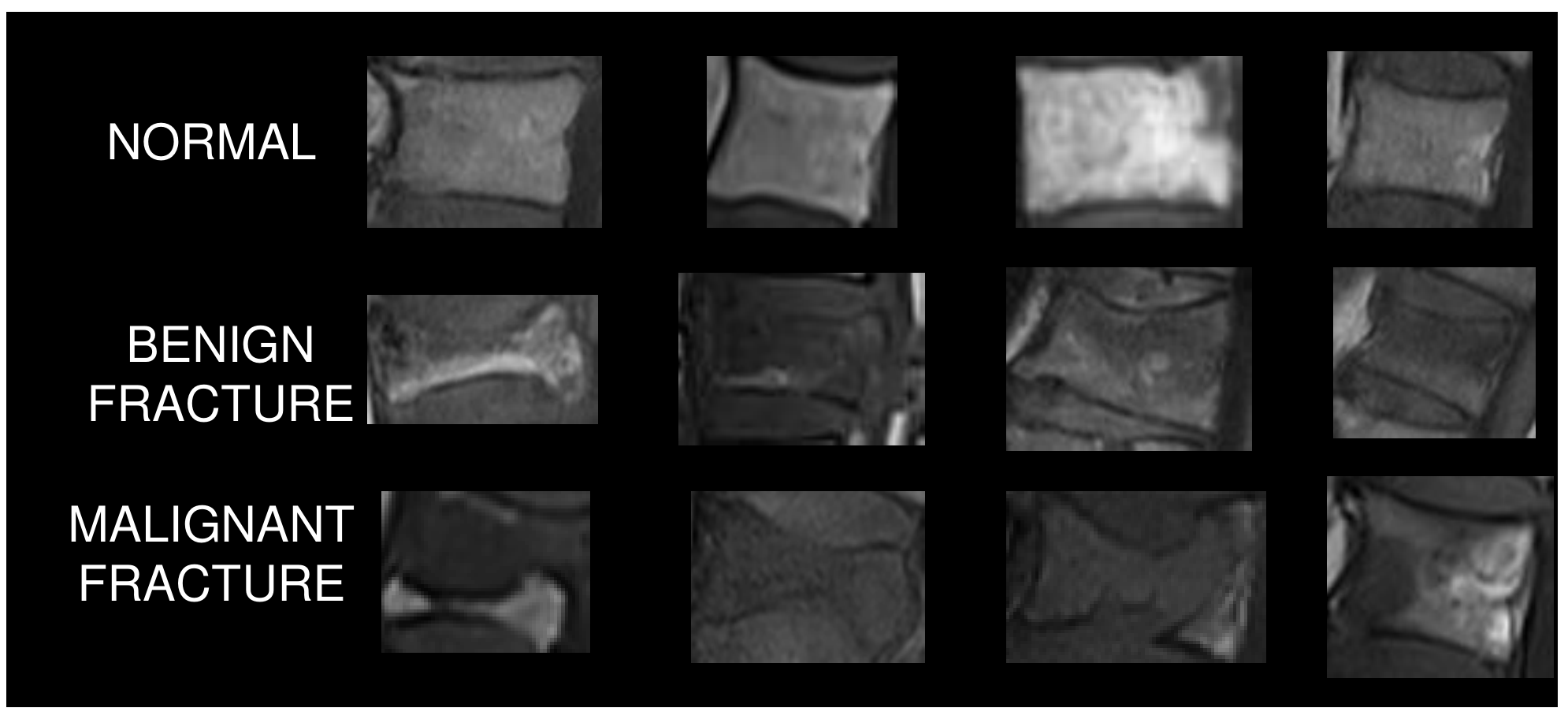}}
\caption{Four examples of each kind of vertebral bodies.}
\label{img3}
\end{figure}

\subsubsection{Evaluation metrics}
As an imbalanced classification task, refer to \cite{wang2018tienet, kim2020m2m}, we calculate sensitivity (SE), specificity (SP), and area under curve (AUC) for each category and use the average number as metrics, i.e., average sensitivity (aSE), average specificity (aSP), and average AUC (aAUC). 
\begin{equation}
\label{eq7a}
\begin{aligned}
aSE&=\frac{1}{K}\sum_{k=1}^K\frac{TP_k}{TP_k+FN_k}, \\
aSP&=\frac{1}{K}\sum_{k=1}^K\frac{TN_k}{TN_k+FP_k}, \\
aAUC&=\frac{1}{K}\sum_{k=1}^K AUC_k,
\end{aligned}
\end{equation}
where $TP_k, FN_k, TN_k, FP_k$ denote number of true positive, false negative, true negative and false positive for class $k$, respectively.
We also adopt mean Average Precision (mAP) as a metric.
\begin{equation}
\label{eq7b}
\begin{aligned}
mAP&=\frac{1}{K}\sum_{k=1}^K AP_k,
\end{aligned}
\end{equation}
where $AP_k$ denote average precision of class $k$. In this paper, $K=3$.

\subsection{Implementation Details}
For our experiments, the slices located at the center of the spine volume is located and used. Specifically, we choose three slices in the middle per patient for training.
WISS and post-processing method introduced in Sec \ref{seg} are applied to segment spine MRI images.
After obtaining the segmentation masks, vertebral bodies are cropped with margin from the slices. Examples of vertebral bodies are shown in Fig. \ref{img3}. 

The dataset is split into $3:1:1$ as training, validation, and test sets. Because the data is imbalanced, we over-sample the minority classes. The patches cropped from MRI slices are resized to $224\times 224$. Rotation, horizontal and vertical flipping are used for data augmentation. 

For all our experiments, ResNet-18 is used as the backbone of the proposed model.
The optimizer is Adam and the batch size is 64. The learning rate is 1e-4 and the weights of $L_{CE},L_D,L_{C}$ and $L_W$ are 1:0.2:1:1. We train our model for 100 epochs and repeat the experiments for 5 times.

\subsection{Methods under Comparison}
According to the characteristics of our dataset, class-imbalanced and fine-grained, we compare with methods belong to imbalanced classification (RS \cite{japkowicz2000class}, SMOTE \cite{chawla2002smote}, CB-RW \cite{cui2019class}, LDAM \cite{cao2019learning}, LDAM + DRW \cite{cao2019learning}) and fine-grained classification (PC-Net \cite{dubey2018pairwise}, NTS-Net \cite{yang2018learning}, SDL-Net \cite{zhang2018skin}).
\begin{itemize}[leftmargin=0.15in]
\item RS \cite{japkowicz2000class} over-samples the two minority classes using different sampling probability for each sample;
\item SMOTE\cite{chawla2002smote} is a variant of re-sampling methods with data augmentation; 
\item Class-balanced re-weighting (CB-RW) \cite{cui2019class} uses the inverse of effective number for each class, defined as $(1-\beta^{N_k})/(1-\beta)$. Here, we use $\beta=0.999$. 
% Deferred re-sampling (DRS) \cite{cao2019learning} applies re-sampling until the later stage of the training. 
\item Label-distribution-aware margin (LDAM)\cite{cao2019learning} proposes a theoretically-principled label-distribution-aware margin (LDAM) loss motivated by minimizing a margin-based generalization bound. LDAM+DRW \cite{cao2019learning} further applies re-weighting until the later stage of the training. 
\item PC-Net \cite{dubey2018pairwise} constructs a Siamese neural network trained with a loss function that attempts to bring class conditional probability distributions closer to each other.
\item NTS-Net \cite{yang2018learning} enables the navigator agent to detect the most informative regions under the guidance from the Teacher agent and the Scrutinizer agent scrutinizes the proposed regions from navigator and makes predictions.
\item SDL-Net \cite{zhang2018skin} uses dual neural networks and enables them to mutually learn from each other by predicting whether the pair of inputs images belong to the same class.
For PC-Net, NTS-Net and SDL-Net, the over-sampling strategy used when training is the same as that used in TSCCN.
% Husseini \textit{et al.} \cite{husseini2020grading} designed grading loss for VCFs recognition and fracture grading which includes a 'ranking' within the different grades of fractures. 

\end{itemize}

\begin{figure}[ht]
\begin{center}
\includegraphics[width=\linewidth]{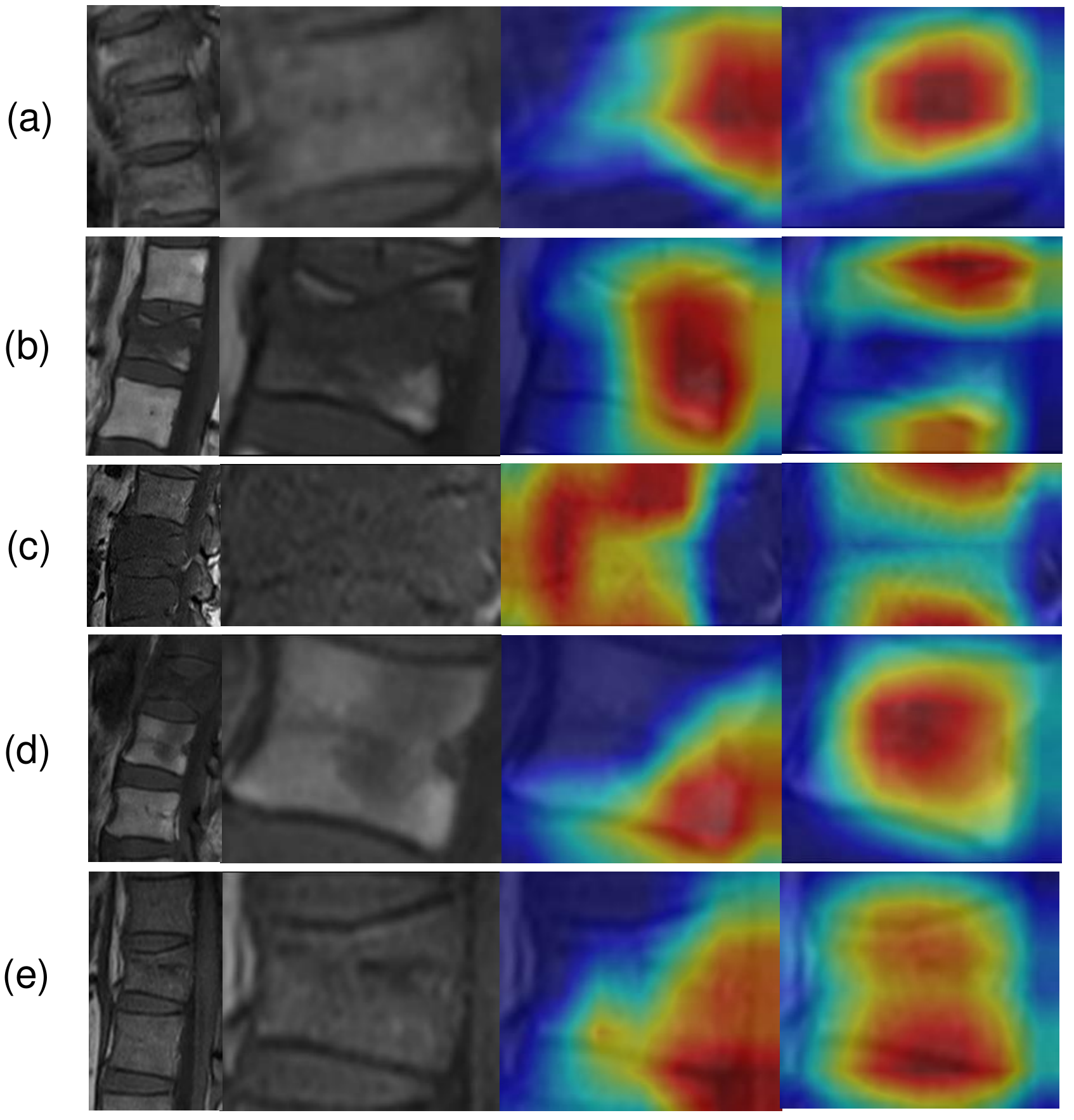}
\end{center}
   \caption{The visualization of CAM for ResNet and TSCCN. The first and second columns show vertebral bodies. The third and last columns show cam images of ResNet and TSCCN, respectively. TSCCN predicted correctly in all the 5 examples, while ResNet predicted wrongly in the last 3 examples. For the 5 examples, the ground truth labels and labels predicted by ResNet are [0, 0], [1, 1], [2, 2], [0, 1], [1, 0], respectively.}
\label{img11}
\end{figure} 

\subsection{Quantitative Results}

% \begin{figure*}[!t]
\begin{figure*}[h]
\begin{center}
    \subfigure[ResNet]{
    \includegraphics[width=0.18\linewidth]{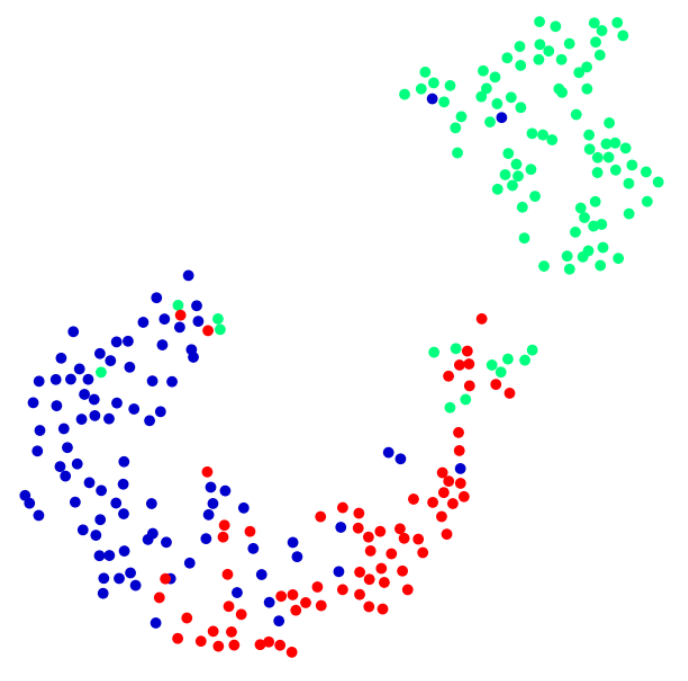}
  }
    \subfigure[PC-Net]{
    \includegraphics[width=0.18\linewidth]{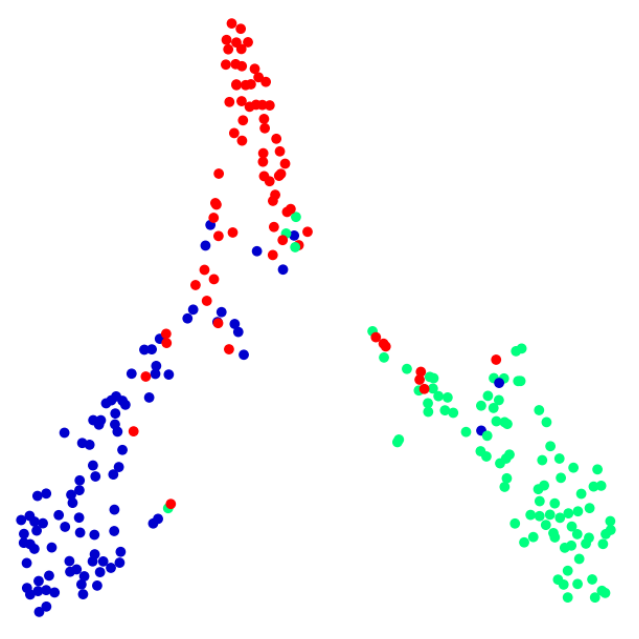}
  }
    \subfigure[NTS-Net]{
    \includegraphics[width=0.18\linewidth]{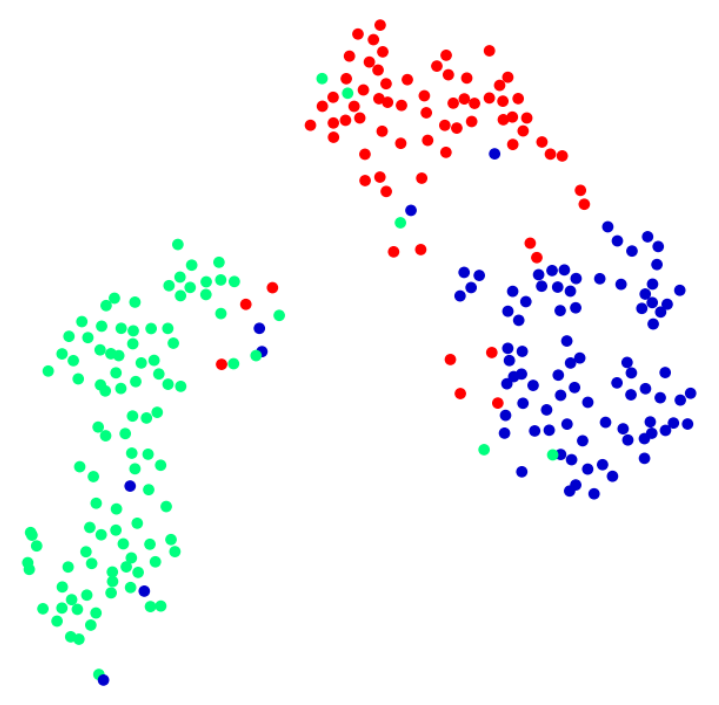}
  }
    \subfigure[SDL-Net]{
    \includegraphics[width=0.18\linewidth]{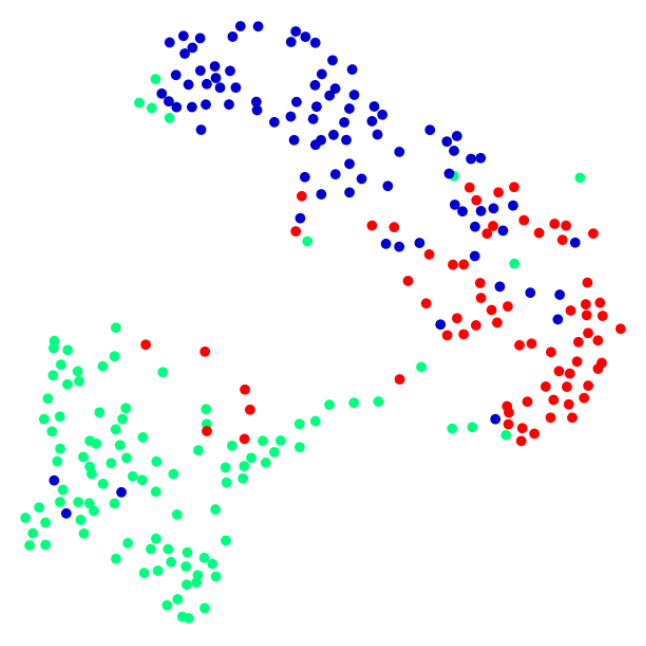}
  }
    \subfigure[RS]{
    \includegraphics[width=0.18\linewidth]{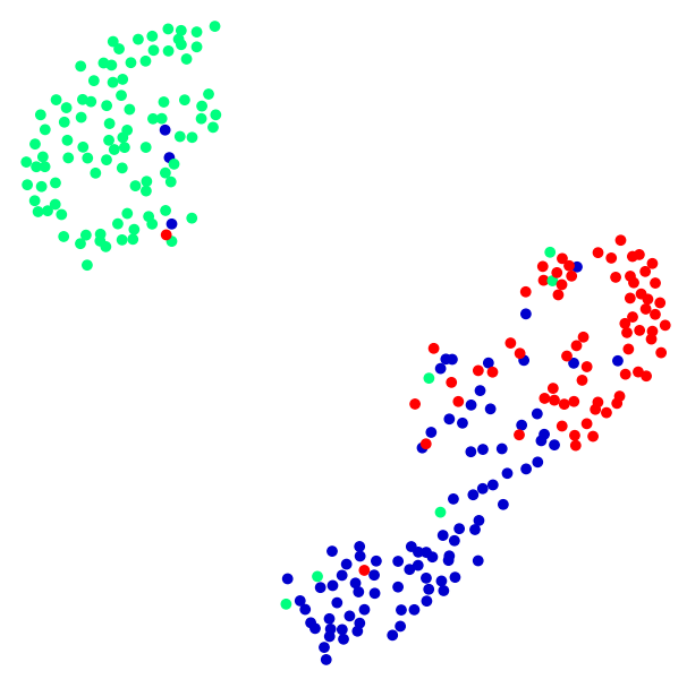}
  }
    \subfigure[CB-RW]{
    \includegraphics[width=0.18\linewidth]{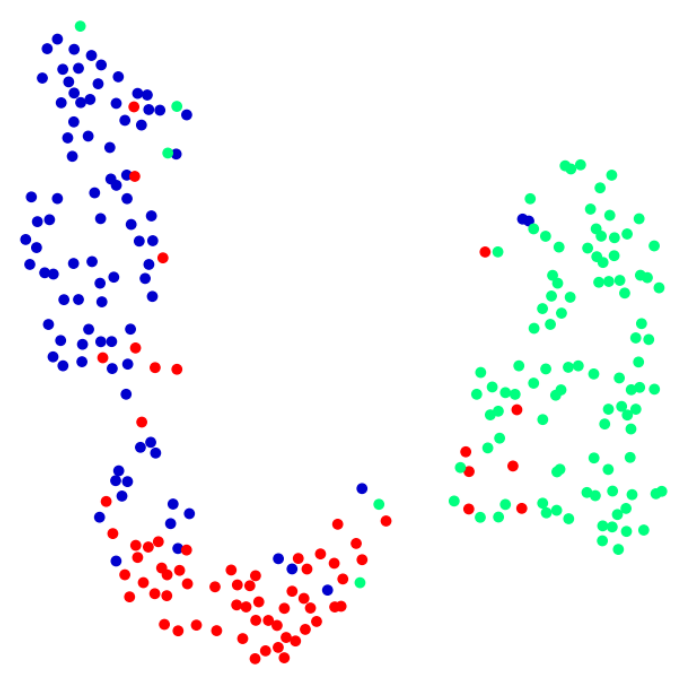}
  }
    \subfigure[SMOTE]{
    \includegraphics[width=0.18\linewidth]{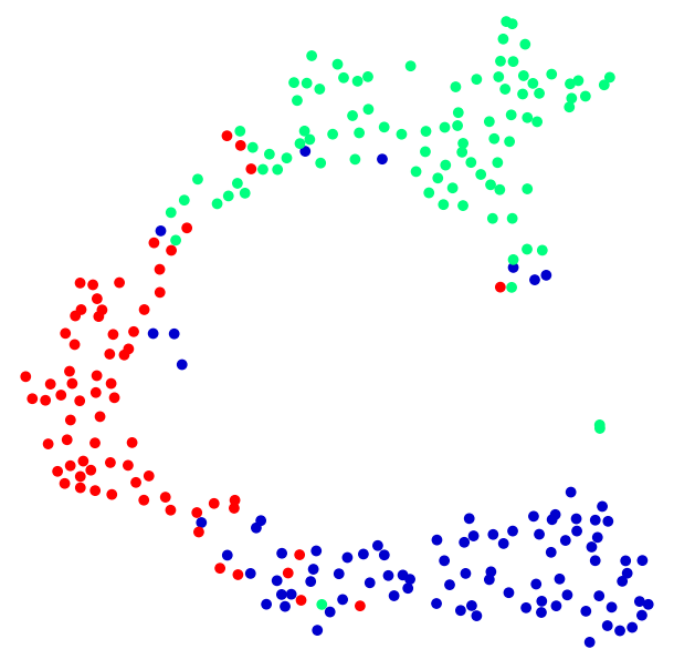}
  }
    \subfigure[LDAM]{
    \includegraphics[width=0.18\linewidth]{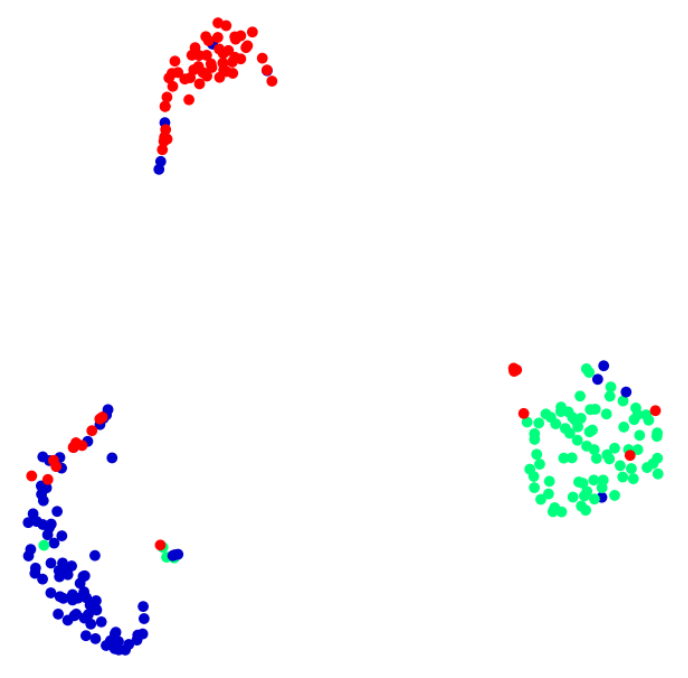}
  }
    \subfigure[LDAM DRW]{
    \includegraphics[width=0.18\linewidth]{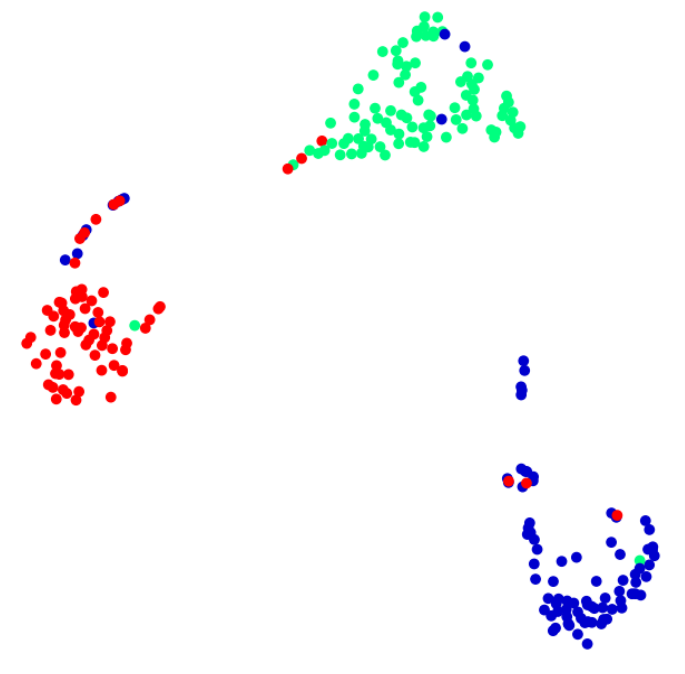}
  }
    \subfigure[Ours]{
    \includegraphics[width=0.18\linewidth]{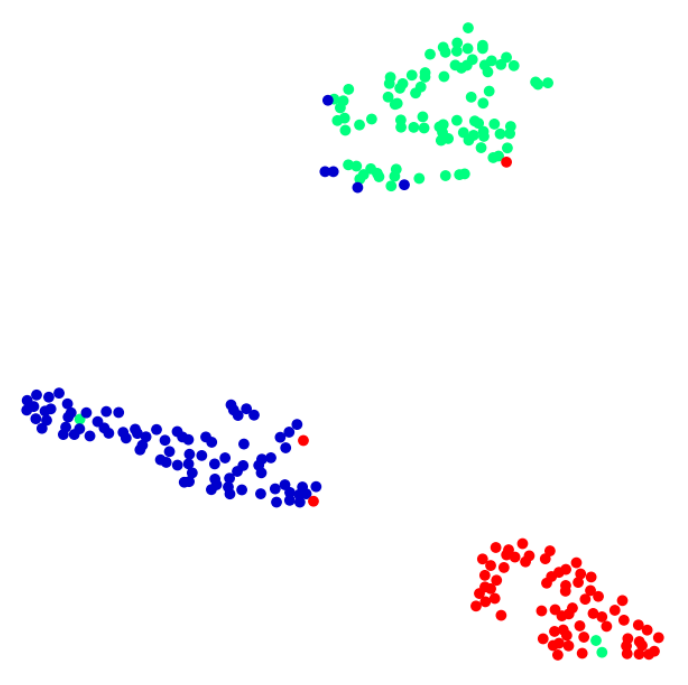}
  }
    
\end{center}
   \caption{The visualization of classification networks via the TSNE algorithm. Green, blue, red are used to denote samples of normal vertebrae, benign VCFs, and malignant VCFs, respectively.}
\label{img6}
\end{figure*}

The experimental results in Table \ref{tab2} have shown that our method leads to the best performance in all metrics. Compared to SDL-Net, which is proposed for skin lesion classification and also faces the problem of imbalanced dataset and fine-grained recognition, the aSE and aSP increase by $2.68\%$ and $1.31\%$ respectively, and the aAUC and mAP increase by $1.35\%$ and $2.14\%$ respectively, which shows the advantages of the network in the overall three-class classification ability. Based on the results for each category, the SE and SP of the two subtypes of VCFs are significantly improved, which demonstrates the effectiveness of TSCCN for improving the performance of minority classes. SE and SP of benign VCFs and SE of malignant VCFs, increase by $0.74\%$ ($91.85\%$ vs $91.11\%$), $1.54\%$ ($96.85\%$ vs $95.31\%$), and $2.42\%$ ($90.71\%$ vs $88.29\%$) respectively, compared to the best performance of all the other methods. And for the SP of malignant VCFs, TSCCN also gets a good performance, with only $0.11\%$ less than NTS-Net.  And for SE and SP of normal vertebrae, we get comparable performance. There is a trade-off relationship between the performance of normal vertebrae and VCFs. Slight performance degradation for normal vertebrae is acceptable in exchange for a huge improvement in performance for VCFs.

\subsection{Qualitative Results}
\subsubsection{CAM}
We show CAM images to compare TSCCN and ResNet. Three consecutive vertebrae are shown to conveniently compare the current vertebra to its adjacent vertebrae. For Fig. \ref{img11} (a), (b), and (c), two networks both predicted correctly, while TSCCN can focus on more reasonable areas. For Fig. \ref{img11} (d), ResNet predicted the normal vertebra as a fractured vertebra since it focus on the right edge while TSCCN predicted it correctly. For Fig. \ref{img11} (e), ResNet predicted it as a normal vertebra while TSCCN predicted correctly since TSCCN focused on the upper and lower edges.

\subsubsection{TSNE}
We visualize the features of FC layer of each model using TSNE, as shown in Fig. \ref{img6}.
Compared with ResNet (Fig. \ref{img6} (a)), methods belong to fine-grained classification (Fig. \ref{img6} (b-d)) have clearer boundary between normal and VCFs and between subtypes of VCFs. Methods belong to imbalanced classification (Fig. \ref{img6} (e-i)) have more equal distance between the three classes. TSCNN achieves both effects. The sample points of  the two subtypes of VCFs in TSNE image of TSCCN are more centralized and are less mixed, and the distance between the three classes are almost equal.

\begin{table}
\caption{Results of models with different amount of data. The same over-sampling strategy is used for the two models. The best result for each column is marked in bold.}
\setlength{\tabcolsep}{3pt}
% \center
\centering
% \begin{tabular}{p{35pt}|p{32pt}|p{30pt}|p{30pt}|p{30pt}|p{30pt}}
\begin{tabular}{|l|c|c|c|c|c|}

\hline
Method & Data size & aSE(\%) & aSP(\%)& aAUC(\%) & mAP(\%) \\
\hline
ResNet & \multirow{2}{*}{Quarter} & 76.60 & 88.51 & 91.77 & 83.94\\
Ours & & 81.65 & 90.87 & 93.95 & 88.65 \\
\hline
ResNet & \multirow{2}{*}{Half} & 82.47 & 91.28 & 93.80 & 88.27\\
Ours & & 86.34 & 93.19 & 95.99 & 92.23 \\
\hline
ResNet & \multirow{2}{*}{Full} & 85.56 & 92.85 & 96.33 & 93.42\\
Ours & & \textbf{92.56} & \textbf{96.29} & \textbf{98.35} & \textbf{97.01} \\

\hline

\end{tabular}
\label{tab6}
\end{table}

\subsection{Generalization across Amount of Data}
We use a quarter, a half, and a full amount of data to train TSCCN and ResNet. As shown in Table \ref{tab6}, for aSE, gains obtained by TSCCN to ResNet18 are $5.05\%$ ($81.65\%$ vs $76.60\%$), $3.87\%$ ($86.34\%$ vs $82.47\%$), and $7.00\%$ ($92.56\%$ vs $85.56\%$), respectively. And we can find the results of our model trained with half of the data is similar to the results of ResNet trained with full of the data, and a similar phenomenon appears when our model is trained with a quarter of the data and ResNet is trained with half of the data. The experimental results demonstrate that although our model has more parameters, its generalization ability is ensured by the well designed two-stream network and embedded expert knowledge which is important for learning distinguishable features from a small amount of data.

\begin{figure}[!t]
\begin{center}
    \subfigure[]{
    \includegraphics[width=0.16\linewidth]{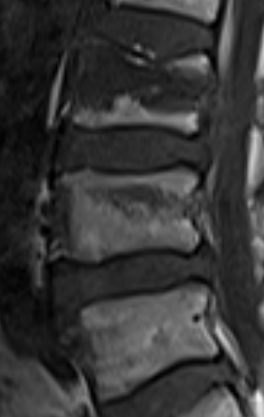}
  }
    \subfigure[]{
    \includegraphics[width=0.16\linewidth]{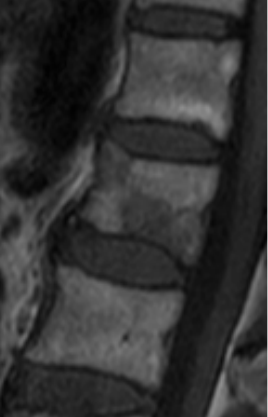}
  }
    \subfigure[]{
    \includegraphics[width=0.16\linewidth]{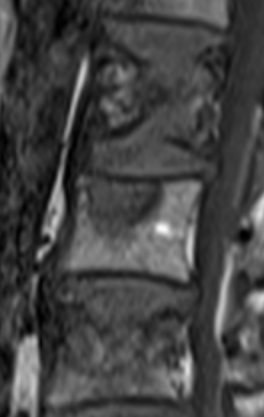}
  }
    \subfigure[]{
    \includegraphics[width=0.16\linewidth]{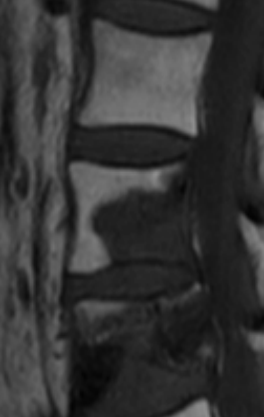}
  }
    \subfigure[]{
    \includegraphics[width=0.16\linewidth]{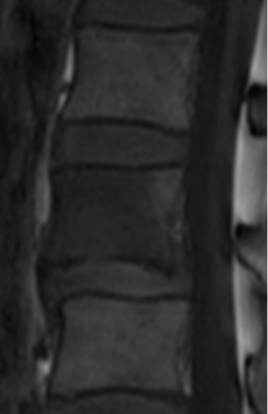}
  }
    
\end{center}
   \caption{Examples of wrongly predicted vertebrae by TSCCN. Each image shows three vertebrae. All the vertebrae in the center are normal. The center vertebrae in (a), (b), (c) are diagnosed as benign VCFs and vertebrae in (d), (e) are diagnosed as malignant VCFs.}
\label{img10}
\end{figure}

\begin{figure*}[h]
\begin{center}
    \subfigure[Single-stream]{
    \includegraphics[height=0.25\linewidth]{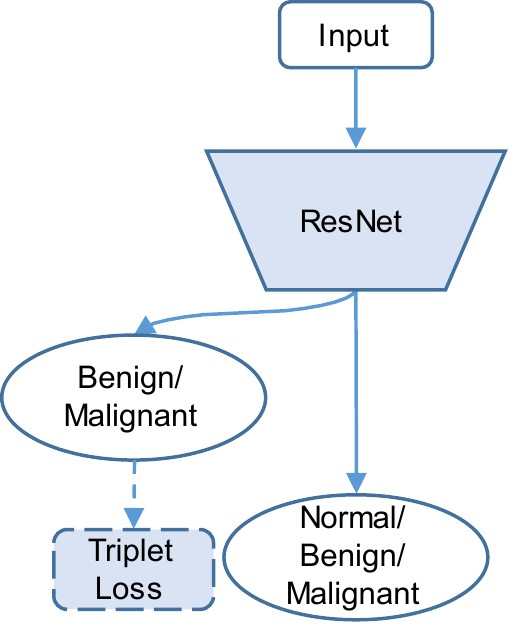}
  }\hspace{3mm}
    \subfigure[Two-stream]{
    \includegraphics[height=0.25\linewidth]{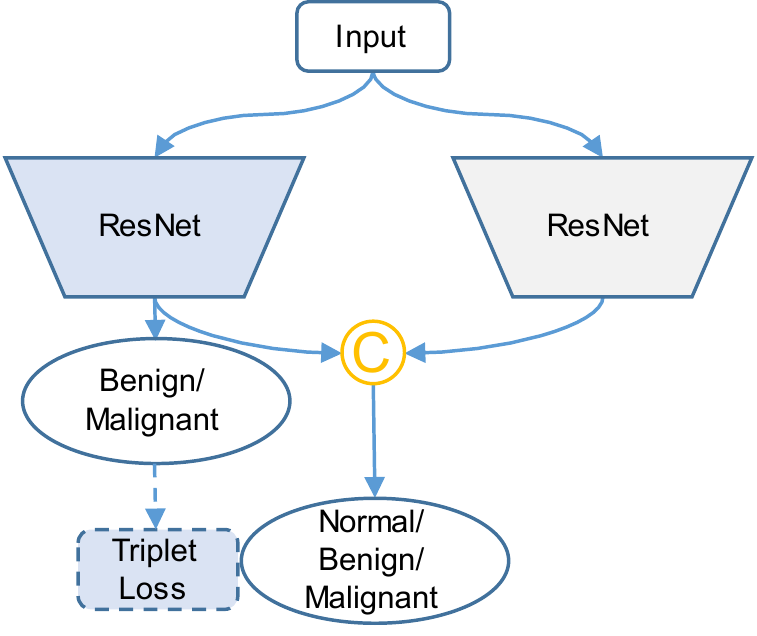}
  }\hspace{3mm}
    \subfigure[Two-stream + compare network]{
    \includegraphics[height=0.25\linewidth]{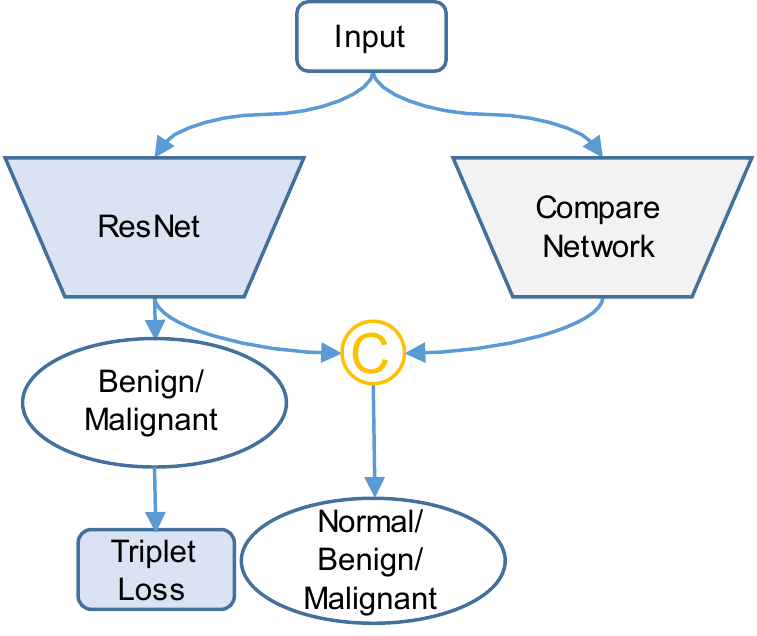}
  }
\end{center}
   \caption{Three network structures for investigating the effectiveness of our two-stream architecture and compare network. Dotted line means triplet loss is optional.}
\label{two-stream}
\end{figure*}

\subsection{Analysis for Wrongly Predicted Vertebrae of TSCCN}
When diagnosing VCFs, there are some deceases which may affect the accuracy of diagnosis.  
We provide some examples of wrongly predicted vertebrae predicted by TSCCN in Fig. \ref{img10}. As shown in Fig. \ref{img10} (a), the normal vertebra is predicted as benign VCFs since its shape and signal intensity are different from its adjacent normal vertebrae. The differences in shape and signal intensity are caused by bone contusion, which is similar to slight fractured. Fig. \ref{img10} (b) and (c) show examples of normal vertebrae being wrongly predicted as benign vertebrae. These two vertebrae suffer from bone degeneration so that their signal intensity is abnormal. Fig. \ref{img10} (d), (e) show examples of normal vertebrae being wrongly predicted as malignant vertebrae. Since malignant VCFs are caused by bone metastasis, there are sometimes not only one vertebra suffering from the invasion of tumor. These two vertebrae are invaded by tumor but they have normal shape. So wrongly predicting these normal vertebrae is not completely meaningless and wrong in clinical. More training data and more accurate ground truth labels are needed to address this problem.

\section{Ablation Study}
We further investigate the individual contribution of the three key components in TSCCN, i.e. the two-stream architecture, the compare network, and the weight control module, via the ablation study.

\subsection{Analysis for Two-Stream Architecture}
We compare the performance of the single-stream architecture (Fig. \ref{two-stream}(a)) and the two-stream architecture (Fig. \ref{two-stream}(b)).
For single-stream network, two classification heads are added to the same encoder. For two-stream network, the features from the two streams are concatenated for three-class classification. Triplet loss is added to binary classification head like the TSCCN. As shown in Table \ref{tab3}, the performance of two-stream network is similar to single-stream network, but two-stream architecture provides the possibility to optimize each stream individually thus easing the trade-off relationship. Compared to single-stream network, the aSE and aSP of two-stream network with triplet loss increase by $1.22\%$ ($88.65\%$ vs $87.43\%$) and $0.61\%$ ($94.36\%$ vs $93.75\%$), respectively.

We notice that adding triplet loss damages the performance of the single-stream network. Triplet loss can enhance the feature extraction ability of single-stream network for benign-malignant specialized features so that weaken the feature extraction ability for normal-fractured specialized features.
Experiments prove our hypothesis that decoupling the two kinds of features using two streams and enhance each stream can improve the three-class classification performance.
 
\begin{table}
\caption{Results of the ablation study. The same over-sampling strategy is used for all the methods. The best result for each column is marked in bold.}
\setlength{\tabcolsep}{3pt}
\center
\begin{tabular}{|p{85pt}|p{30pt}|p{30pt}|p{30pt}|p{30pt}|}
% \begin{tabular}{lcccc}

\hline
Method & aSE(\%) & aSP(\%)& aAUC(\%) & mAP(\%) \\
\hline
single-stream & 87.43 & 93.75 & 96.53 & 93.90\\
\hline
single-stream + triplet loss & 86.89 & 93.49 & 96.43 & 93.68 \\
\hline
% two-stream-share & 86.75 & 93.41 & 96.13 & 92.99 \\
two-stream & 87.43 & 93.95 & 96.38 & 93.49 \\
\hline
two-stream + triplet loss & 88.65 & 94.36 & 96.80 & 94.39\\
\hline
two-stream + triplet loss + compare network& \multirow{2}{*}{91.12} & \multirow{2}{*}{95.56} & \multirow{2}{*}{98.28} & \multirow{2}{*}{96.91} \\
\hline
TSCCN & \textbf{92.56} & \textbf{96.29} & \textbf{98.35} & \textbf{97.01} \\

\hline
\end{tabular}
\label{tab3}
\end{table}

\begin{figure}[!t]
\begin{center}
    \subfigure[TSNE for Single-stream]{
    \includegraphics[width=0.85\linewidth]{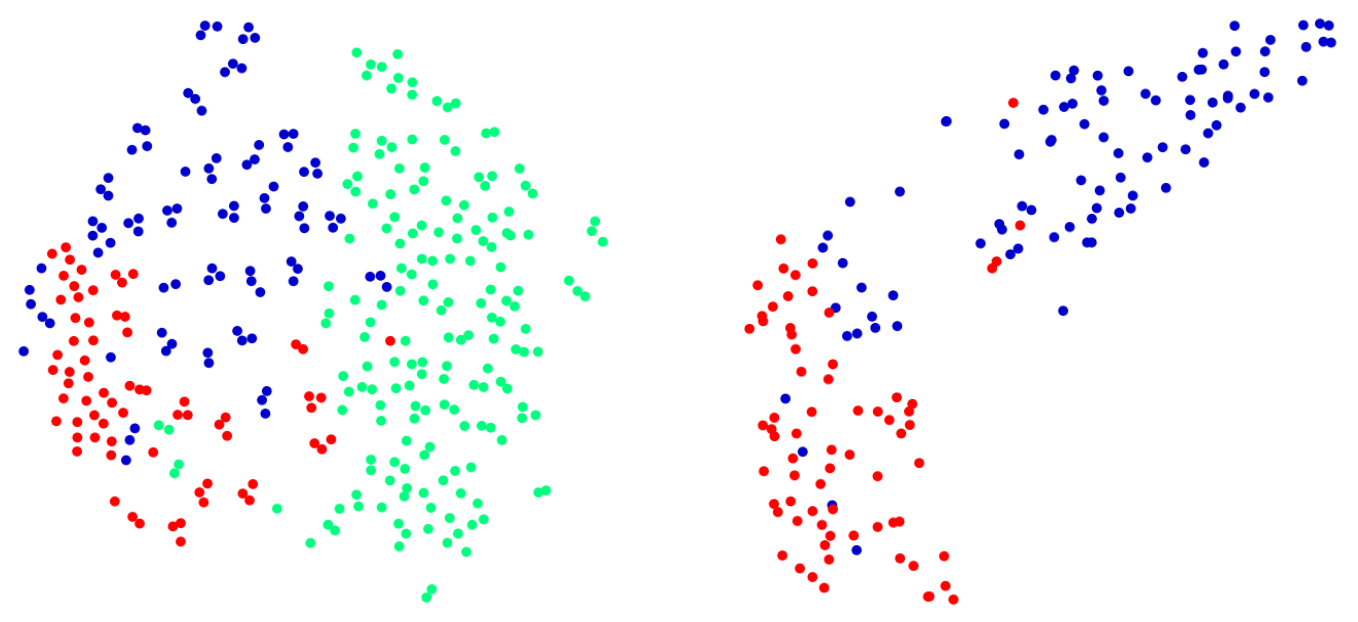}
  }
    \subfigure[TSNE for Two-stream]{
    \includegraphics[width=0.85\linewidth]{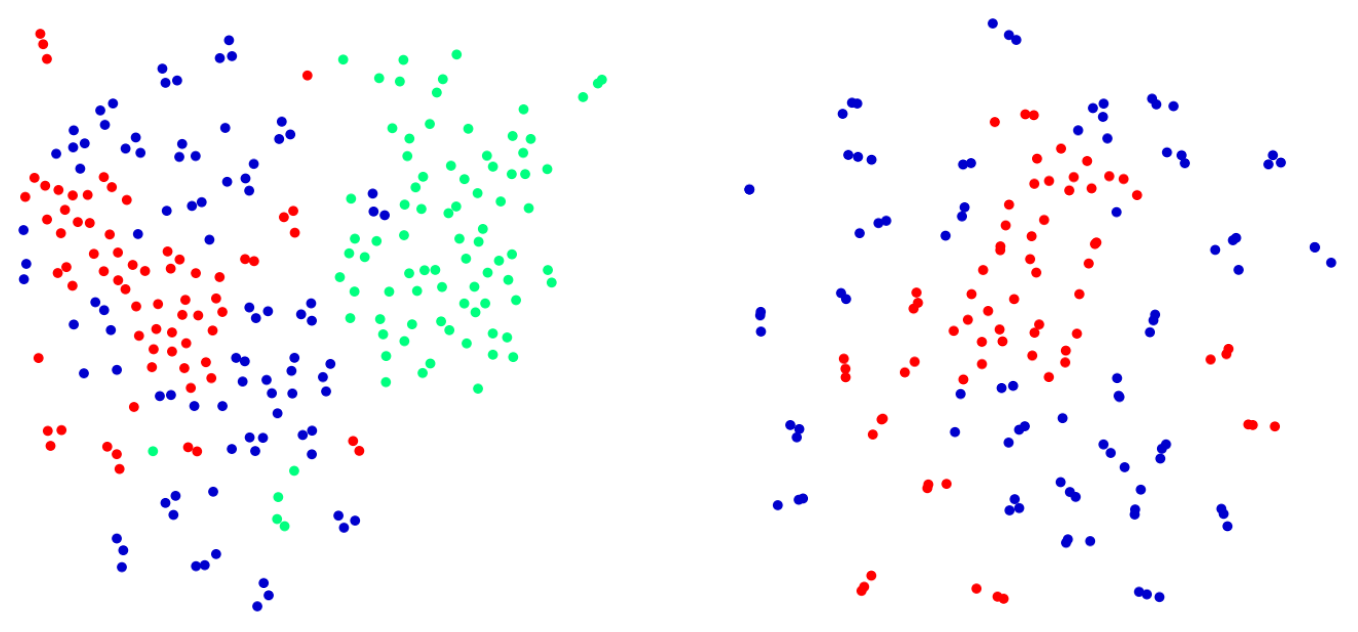}
    }
    \subfigure[TSNE for Two-stream with compare network]{
    \includegraphics[width=0.85\linewidth]{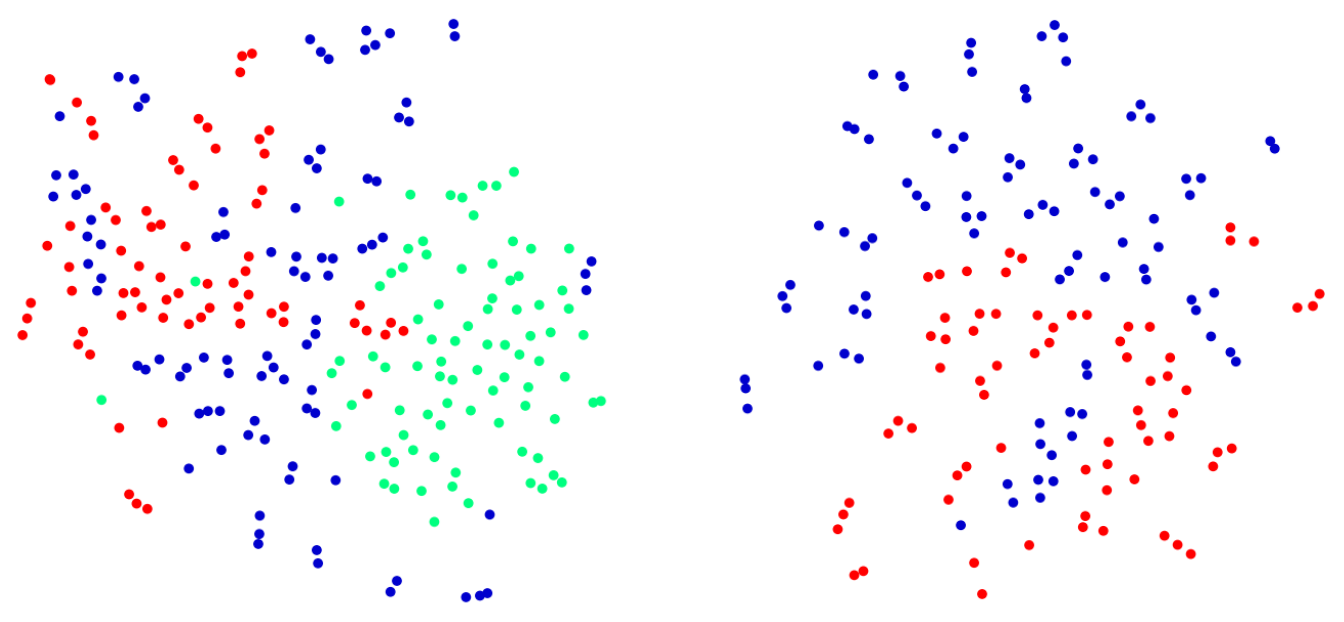}
    }
    % \subfigure[T-SNE for TSCCN]{
    % \includegraphics[width=0.9\linewidth]{img/l3/context_concat_triplet_controlWR4_l3.pdf}
    % }
\end{center}
  \caption{Illustration of two-stream architecture and compare network. For (a), features are obtained from the 3rd res-block of single-stream, and for (b) and (c), features are from recognition stream of two-stream network. The second column show the two-class (benign-malignant) TSNE images to better show the mixture degree of them. Green, blue, red are used to denote samples of normal vertebrae, benign VCFs, and malignant VCFs, respectively.}
\label{img7}
\end{figure}

Fig. \ref{img7}(a) and (b) shows the TSNE images for single-stream and two-stream network. The features of encoder of single-stream network presents clustering distribution for each category, as shown in Fig. \ref{img7}(a). For the features of recognition stream of two-stream network, the samples of normal and VCFs are correctly separated, while samples of benign and malignant VCFs mix together, as shown in Fig. \ref{img7} (b). It shows that recognition stream extracts features only for differentiating between normal vertebrae and VCFs.

\subsection{Analysis for Compare Network}
As shown in Table \ref{tab3}, the proposed compare network greatly improves the overall classification performance.
The aSE and aSP are larger than those of the two-stream network without compare network by $2.47\%$ ($91.12\%$ vs $88.65\%$) and $1.20\%$ ($95.56\%$ vs $94.36\%$) respectively.
The big promotion of classification performance is caused by that compare network can enable the network to compare and contrast between adjacent vertebrae so that can learn contextual features better.
As shown in Fig. \ref{img7}(c), for the benign and malignant VCFs, the mixture degree of features from the recognition stream is deeper, which means the compare network learns more specialized features for differentiating between normal vertebrae and VCFs, compared to those from the recognition stream of two-stream network, as shown in Fig. \ref{img7}(b).

\begin{figure}[!t]
\begin{center}
    \includegraphics[width=0.65\linewidth]{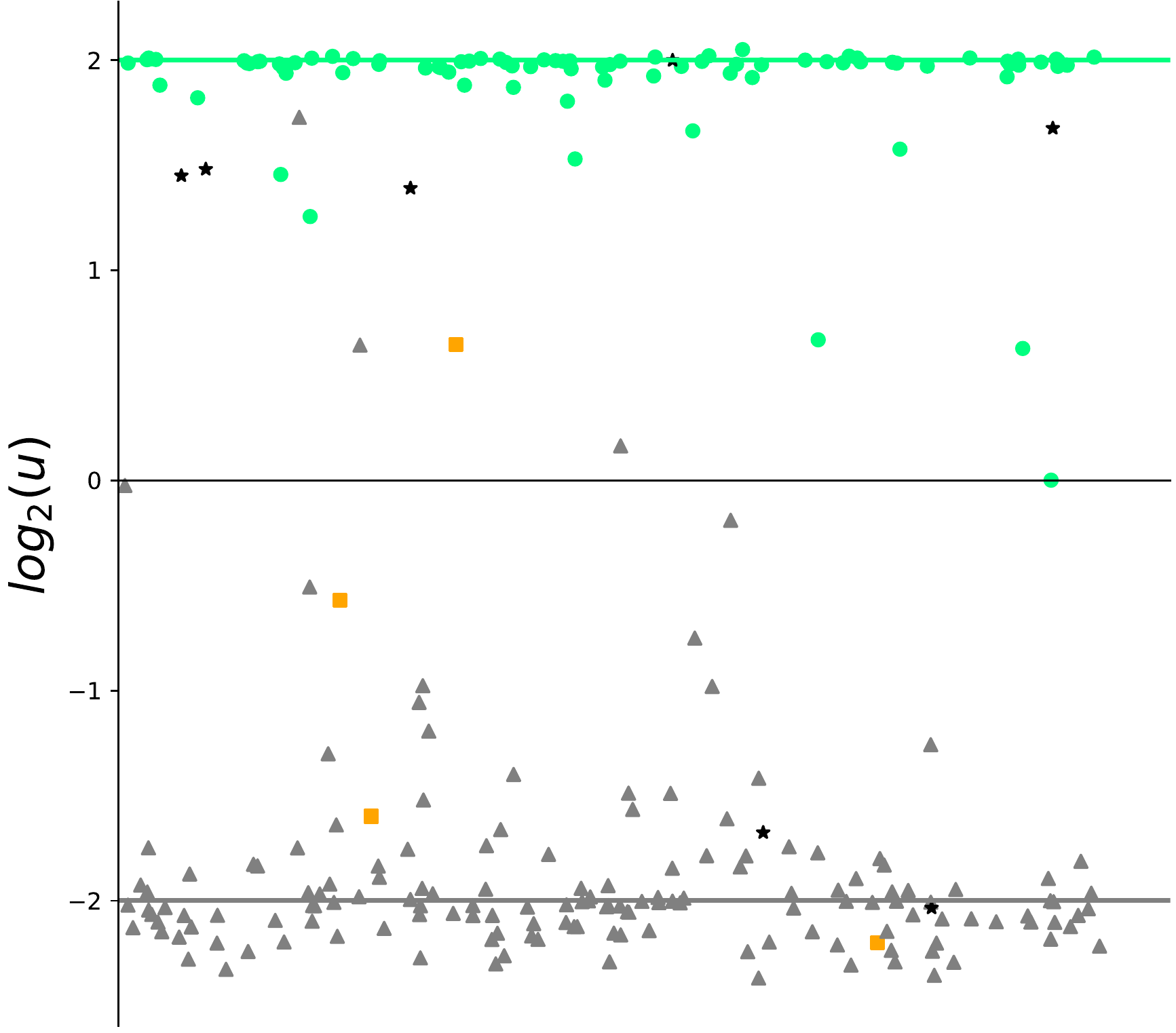}
\end{center}
  \caption{An example of relation between weight ratio $u$ and ground truth label. Green circle and gray triangle are used to denote correctly predicted samples of normal vertebrae and VCFs, respectively. Orange square and black stars are used to denote wrongly predicted samples of normal vertebrae and VCFs, respectively.}
\label{img9}
\end{figure}

\subsection{Analysis for Weight Control Module}
The weight control module is designed to emphasize the differences of output features between the two streams based on our hypothesis that two streams learn differently biased features and matters differently according to the input class. As shown in Table \ref{tab3}, the weight control module further boost the performance of the two-stream network. The aSE and aSP increase by $1.44\%$ ($92.56\%$ vs $91.12\%$) and $0.73\%$ ($96.29\%$ vs $95.56\%$), respectively. 
To better understand this module, we show the relationship between the value $u$ of weight control module and the ground truth label. As we can see in Fig. \ref{img9}, the green circles which denote the correctly predicted normal vertebrae are almost near the value 2. $u$ of the green circles are larger than 1, which means when predicting the label of these vertebrae the recognition stream is more important. And $u$ of almost all of the orange squares which denote the wrongly predicted normal vertebrae are smaller than 1. And the similar phenomenon occurs for fractured vertebrae too.

\section{Conclusion}
\label{sec5}
In this work, the VCFs recognition and VCFs classification tasks are combined as a three-class classification task. We propose a novel Two-Stream Compare and Contrast Network (TSCCN) which uses two streams to individually enhance feature extracting ability for recognition and classification and a weight control module to better integrate features from two-streams.
We point out that comparing and contrasting  between adjacent vertebrae utilizing the continuity of spine and comparing and contrasting between benign and malignant vertebrae are helpful for improving the accuracy of vertebrae diagnosis. 
We demonstrate our model in our vertebrae diagnosis dataset and surpass the previous methods.

{\small
\bibliographystyle{ieeetr}
\bibliography{tmi}
}

\end{document}